\documentclass[dvips]{article}
\usepackage{icrctc07}

\title{H.E.S.S. observations of the supernova remnant RX~J0852.0-4622: 
shell-type morphology and spectrum of a widely extended VHE gamma-ray 
source}
\shorttitle{H.E.S.S. observations of the SNR RX~J0852.0-4622}

\authors{M. Lemoine-Goumard$^{1}$, F. Aharonian$^{2,3}$, B. Degrange$^{4}$, L. Drury$^{2}$, U. Schwanke$^{5}$, for the H.E.S.S. Collaboration}
\shortauthors{M. Lemoine-Goumard (for the H.E.S.S. Collaboration) et al.}
\afiliations{$^1$CENBG, Universit\'e Bordeaux I, CNRS-IN2P3, Le Haut-Vigneau, 33175 Gradignan, France\\ $^2$Dublin Institute for Advanced Studies, 5 Merrion Square, Dublin, Ireland\\$^3$Max-Planck-Institut f\"ur Kernphysik, P.O. Box 103980, Heidelberg, Germany\\$^4$LLR, Ecole Polytechnique, CNRS-IN2P3, 91128 Palaiseau, France\\$^5$Institut f\"ur Physik, Humboldt-Universit\"at zu Berlin, D 12489 Berlin, Germany}
\email{lemoine@cenbg.in2p3.fr}

\abstract{The shell-type supernova remnant RX~J0852.0-4622 was detected in 2004 and re-observed between December 2004 and May 2005 with the High Energy Stereoscopic System (H.E.S.S.), an array of four Imaging Cherenkov Telescopes located in Namibia and dedicated to the observations of $\gamma$-rays above 100~GeV. The angular resolution of $< 0.1^{\circ}$ and the large field of view of H.E.S.S. ($5^{\circ}$ diameter) are well adapted to studying the morphology of the object in very high energy gamma-rays, which exhibits a remarkably thin shell very similar to the features observed in the radio range and in X-rays. The spectral analysis of the source from 300~GeV to 20~TeV will be presented. Finally, the possible origins of the very high energy gamma-ray emission (Inverse Compton scattering by electrons or the decay of neutral pions produced by proton interactions) will be discussed, on the basis of morphological and spectral features obtained at different wavelengths.}

\begin{document}
\maketitle
\section{Introduction}

Shell-type supernova remnants (SNR) are widely believed to be the prime candidates for accelerating cosmic rays up to $10^{15}$~eV, but until recently, this statement was only supported by indirect evidence, namely non-thermal X-ray emission interpreted as synchrotron radiation from very high energy electrons from a few objects. A more direct proof is provided by the detection of very high energy $\gamma$-rays, produced in nucleonic interactions with ambient matter or by inverse Compton scattering of accelerated electrons off ambient photons.\\
Here, we present recent data on RX~J0852.0$-$4622 obtained with H.E.S.S. in 2004 and 2005. 

\section{The H.E.S.S. detector and the analysis technique}

H.E.S.S. is an array of four 13 m diameter imaging Cherenkov telescopes 
located in the Khomas Highlands in Namibia, 1800~m above sea level~\cite{HESS}. Each telescope has a tesselated mirror with an area of 107~m$^{2}$~\cite{HESSOptics}
and is equipped with a camera comprising 960 photomultipliers~\cite{HESSCamera} 
covering a field of view of 5$^{\circ}$ diameter. Due to the 
powerful rejection of hadronic showers provided 
by stereoscopy, the complete system (operational since December 2003) 
can detect point sources at flux levels of about 1\% of 
the Crab nebula flux near zenith with a significance of 5~$\sigma$ in 25 hours of observation. 
This high sensitivity, the angular resolution of a few arc minutes and the large field of 
view make H.E.S.S. ideally suited for 
the study of the $\gamma$-ray morphology of extended sources. 
During the observations, an array level 
hardware trigger required each shower to be observed by at least two telescopes within a coincidence window of 60~ns~\cite{HESSTrigger}. The data were recorded in runs of typical 28 minute 
duration in the so-called ``wobble mode'', where the source is offset from the center of 
the field of view, and were calibrated as described in detail in~\cite{HESSCalib}. In a first stage, 
a standard image cleaning was applied to shower images to remove the pollution due to 
the night sky background. The results presented in 
this paper were obtained using a 3D-modeling of the light-emitting region of an 
electromagnetic air shower, a method referred to as ``the 3D-model analysis''~\cite{HESSModel3D}, and were also cross-checked with the standard H.E.S.S. stereoscopic analysis based on the Hillas parameters of showers images~\cite{aharonian04}. The excess skymap was produced with a background subtraction called the ``Weighting Method''~\cite{WeightingMethod}. In this method, the signal and 
the background are estimated simultaneously in the same portion of the sky. In each sky bin 
(treated independently), the signal and the background are estimated from those events 
originating from this bin exclusively; this is done by means of a likelihood fit 
in which each event is characterized by a discriminating parameter whose distribution is fairly different for $\gamma$-rays and hadrons. In the case of the 3D-Model, this discriminating parameter is the 3D-width of the electromagnetic shower. 

\section{H.E.S.S. results}
\begin{figure}[htbp]
\centering
\includegraphics[width=0.45\textwidth]{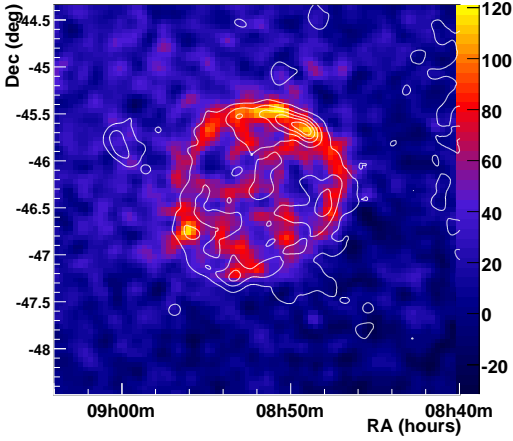}
\caption{Excess skymap of RX~J0852.0$-$4622 smoothed with a Gaussian of 0.06$^\circ$ 
standard deviation, obtained with the 3D-Model. The white lines are the contours of the X-ray data from the 
ROSAT All Sky Survey for energies higher than 1.3 keV (smoothed with a Gaussian 
of 0.06$^\circ$ standard deviation to enable direct comparison of the two images).
\label{fig:velajrmap}}
\end{figure}

\begin{figure}[htbp]
\centering
\includegraphics[width=0.45\textwidth]{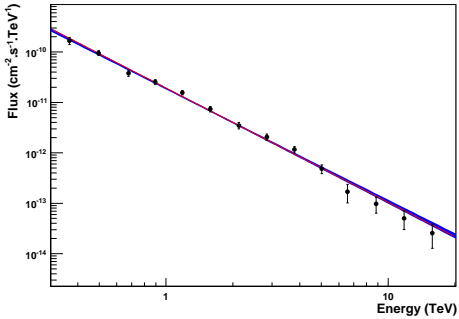}
\caption{Differential energy spectrum of RX~J0852.0$-$4622, for the whole region of 
the SNR. The shaded area gives the 1$\sigma$ confidence region for the spectral 
shape under the assumption of a power law. The spectrum ranges from 300 GeV to 20 TeV.
\label{fig:velajrspec}}   
\end{figure}

RX~J0852.0$-$4622 is a shell-type SNR discovered in the ROSAT all-sky survey. Its X-ray emission is mostly non-thermal~\cite{aschen98}. Indeed, up to now no thermal X-rays were detected from this source, which could imply a limit on the density of the material in the remnant $n_0 < 2.9 \times 10^{-2} (D/ 1 \, \mathrm{kpc})^{-1/2} f^{-1/2} \, \mathrm{cm^{-3}}$, where $f$ is the filling factor of a sphere taken as the emitting volume in the region chosen~\cite{slane}. The X-ray non-thermal spectrum of the whole remnant in the 2-10 keV energy band is well described by a power law with a spectral index of $2.7 \pm 0.2$ and a flux $F_X = 13.8 \times 10^{-11} \, \mathrm{erg \, cm^{-2} \, s^{-1}}$~\cite{velajrApJ}. In the TeV range, the announcement of a signal from the North-Western part of the remnant by CANGAROO was rapidly followed by the publication of a complete $\gamma$-ray map by H.E.S.S. obtained from a short period of observation (3.2~hours)~\cite{HESSVelaJr}. The study of this source is really complex due to several points: its extension (it is the largest extended source ever detected by a Cherenkov telescope), its location at the South-Eastern corner of the Vela remnant and the uncertainty on its distance and age. Indeed, RX~J0852.0$-$4622 could be as close as Vela ($\sim$ 250~pc) and possibly in
interaction with Vela, or as far as the Vela Molecular Ridge ($\sim$
1~kpc). Figure~\ref{fig:velajrmap} presents the $\gamma$-ray image of RX~J0852.0$-$4622 obtained with the 3D-Model from a long observation in 2005 (corresponding to 20~hours live time). The morphology appearing from this skymap reveals a very thin shell of $1^{\circ}$ radius and thickness smaller than $0.22^{\circ}$. Another interesting feature is the remarkably circular shape of this shell, even if the Southern part shows a more diffuse emission. Keeping all events inside a radius of $1^{\circ}$ around the center of the remnant, the cumulative significance is about 19$\sigma$ and the cumulative excess is $\sim 5200$ events. The overall $\gamma$-ray morphology seems to be similar to the one seen 
in the X-ray band, especially in the Northern part of the remnant where a brightening is seen in both wavebands. The correlation coefficient between the $\gamma$-ray 
counts and the X-ray counts in bins of 0.2$^\circ$ $\times$ 0.2$^\circ$ is found 
to be equal to 0.60 and comprised between 0.54 and 0.67 at 95$\%$ confidence level. The differential energy spectrum (Fig.~\ref{fig:velajrspec}) extends from 300~GeV up to 20~TeV. The spectral parameters were obtained from a maximum likelihood fit of a power law 
hypothesis dN/dE = $\mathrm{N_0}$~$\mathrm{(E/1 \, TeV)^{-\Gamma}}$ to the data, resulting 
in an integral flux above 1 TeV of ($15.2 \pm 0.7_{\mathrm{stat}} \pm 3.20_{\mathrm{syst}}$) $\times$ $10^{-12} \mathrm{cm^{-2}} 
\mathrm{s^{-1}}$ and a spectral index of 2.24 $\pm 0.04_{\mathrm{ stat}} 
\pm 0.15_{\mathrm{ syst}}$. An indication of curvature at high energy can be noticed. 

\section{Emission processes}
One of the key issues is the interpretation of the $\gamma$-ray signal in terms of an electronic or a hadronic scenario. Despite the large uncertainty on the distance and age of the remnant, the multi-wavelength data already give some strong constraints. In a leptonic scenario, where $\gamma$-rays are produced by Inverse Compton scattering of high energy electrons off ambient photons, the ratio of the X-ray flux and the $\gamma$-ray flux determines the magnetic field to be close to $7 \, \mu \rm G$. This value is completely independent of the distance and only assumes a filling factor (fraction of the Inverse Compton emitting electrons containing the magnetic field responsible for the synchrotron emission) of 1; this low magnetic field seems hardly compatible with the amplification suggested by the thin filaments resolved by Chandra~\cite{chandra}. In the nearby case ($\sim 200$~pc), the limit on the width of the shell $\Delta R$ obtained by the morphological analysis of the H.E.S.S. data is $\Delta R < 0.7$~pc, which leads to an escape time by diffusion and by convection lower than both the age of the remnant and the synchrotron cooling time for energies higher than $\sim 10$~TeV. Therefore, one would expect to see a variation of the width of the shell with the energy, which is not observed by H.E.S.S.~\cite{velajrApJ} and disfavours the electronic scenario at this distance. In a hadronic scenario, in which we assume that the $\gamma$-ray flux is entirely due to proton-proton interactions, one can estimate the total energy in accelerated protons in the range $10 - 100$~TeV required to produce the $\gamma$-ray luminosity observed by H.E.S.S.:
\begin{eqnarray}
L_{\gamma} (1-10 \, \mathrm{TeV}) & = & 4 \pi D^2 \int_{1\,\mathrm{TeV}}^{10\,\mathrm{TeV}} E \phi(E) dE \nonumber \\ 
                               & = & 2.6 \times 10^{32} \big(\frac{D}{200 \, \mathrm{pc}} \big)^2 \mathrm{erg \, s^{-1}} \nonumber 
\end{eqnarray}
In this energy range, the characteristic cooling time of protons through the $\pi^0$ production channel is approximately independent of the energy and can be estimated to be: $\tau_{\gamma} = 4.4 \times 10^{15} \left(\frac{n}{1 \, \mathrm{cm^{-3}}} \right)^{-1}$. Assuming that the proton spectrum continues down to $E \approx 1$~GeV with the same spectral slope as that of the photon spectrum, the total energy injected into protons is estimated to be:
$$ W_p^{\rm tot} \approx 10^{49} \left( \frac{D}{200 \, \mathrm{pc}} \right)^2  \left( \frac{n}{1 \, \mathrm{cm^{-3}}} \right)^{-1} \mathrm{erg}$$ 
Therefore, for densities compatible with the absence of thermal X-rays, the only way to explain the entire $\gamma$-ray flux by proton-proton interactions in a homogeneous medium is to assume that RX~J0852.0$-$4622 is a nearby supernova remnant (D $< 600$~pc). Indeed, for larger distances and a typical energy of the supernova explosion of $10^{51}$~erg, the acceleration efficiency would be excessive. Nevertheless, a distance of 1~kpc should also be considered if one assumes that RX~J0852.0$-$4622 is the result of a core collapse supernova which exploded inside a bubble created by the wind of a massive progenitor star~\cite{bervolk06}. According to stellar wind theory, the size of the bubble evolves according to the formula: $R = 45 \left(\frac{n_0}{1 \, \rm{cm^{-3}}}\right)^{-0.2}$~pc. For a density of 1~$\mathrm{cm^{-3}}$, the radius of this bubble would be equal to 45~pc. In the case of a close supernova remnant, its size would be significantly lower than the size of the bubble and the hypothesis of a homogeneous medium would be satisfactory. In the opposite, for larger distances ($D \sim 1$~kpc), the presence of the Vela Molecular Ridge can produce a sudden increase of the density leading to a smaller bubble ($15.6$~pc for a density of $200 \, \mathrm{cm^{-3}}$), which would make the proton-proton interactions efficient at the outer shock.
 
\section{Summary}
We have firmly established that the shell-type supernova remnant RX~J0852.0$-$4622 is a TeV emitter and for the first time we have resolved its morphology in the $\gamma$-ray range, which is highly correlated with the emission observed in X-rays. Its overall $\gamma$-ray energy spectrum extends over two orders of magnitude, providing the direct proof that particles of $\sim 100$~TeV are accelerated at the shock.\\
The question of the nature of the particles producing the $\gamma$-ray signal observed by H.E.S.S. was also addressed. In the case of a close remnant, the results of the morphological study combined with our spectral modeling highly disfavour the leptonic scenario which is unable to reproduce the thin shell observed by H.E.S.S. and the thin filaments resolved by Chandra. In the case of a medium distance, the explosion energy needed to explain the $\gamma$-ray flux observed by H.E.S.S., taking into account the limit on the density implied by the absence of thermal X-rays, would highly disfavour the hadronic process. At larger distances, both the leptonic and the hadronic scenario are possible, at the expense, for the leptonic process, of a low magnetic field of $\approx 7 \, \mu \rm G$. Such a small magnetic field exceeds typical interstellar values only slightly and is difficult to reconcile with the theory of magnetic field amplification at the region of the shock.\\
However, at present, no firm conclusions can be drawn from the spectral shape. The results which should hopefully be obtained by GLAST or H.E.S.S. II at lower energies will therefore have a great interest for the domain.

\section{Acknowledgements}
The support of the Namibian authorities and of the University of Namibia
in facilitating the construction and operation of H.E.S.S. is gratefully
acknowledged, as is the support by the German Ministry for Education and
Research (BMBF), the Max Planck Society, the French Ministry for Research,
the CNRS-IN2P3 and the Astroparticle Interdisciplinary Programme of the
CNRS, the U.K. Science and Technology Facilities Council (STFC),
the IPNP of the Charles University, the Polish Ministry of Science and 
Higher Education, the South African Department of
Science and Technology and National Research Foundation, and by the
University of Namibia. We appreciate the excellent work of the technical
support staff in Berlin, Durham, Hamburg, Heidelberg, Palaiseau, Paris,
Saclay, and in Namibia in the construction and operation of the
equipment.

\bibliography{icrc0504}

\begin{thebibliography}{10}

\bibitem{aharonian04}
F.~{Aharonian et al. ({\it H.E.S.S. Collaboration})}.
\newblock {\em APh}, 22:109, 2004.

\bibitem{HESSVelaJr}
F.~{Aharonian et al. ({\it H.E.S.S. Collaboration})}.
\newblock {\em A\&A}, 436:L7, 2005.

\bibitem{velajrApJ}
F.~{Aharonian et al. ({\it H.E.S.S. Collaboration})}.
\newblock {\em ApJ}, 661:236--249, 2007.

\bibitem{HESSCalib}
F.~{Aharonian ({\it H.E.S.S. Collaboration})}.
\newblock {\em APh}, 22:109, 2004.

\bibitem{aschen98}
B.~{Aschenbach et al.}
\newblock {\em Nature}, 396:141, 1998.

\bibitem{chandra}
A.~{Bamba}, R.~{Yamazaki}, and J.S. {Hiraga}.
\newblock {\em ApJ}, 632:294, 2005.

\bibitem{bervolk06}
E.~G. {Berezhko} and H.~J. {V\"olk}.
\newblock {\em A\&A}, 451:981, 2006.

\bibitem{HESSOptics}
K.~{Bernl\"ohr et al.}
\newblock {\em APh}, 20:111, 2003.

\bibitem{HESSTrigger}
S.~{Funk et al.}
\newblock {\em APh}, 22:285, 2004.

\bibitem{HESS}
J.~A. {Hinton}.
\newblock {\em NewAR}, 48:331, 2004.

\bibitem{WeightingMethod}
M.~{Lemoine-Goumard} and B.~{Degrange}.
\newblock In {\em Proceedings of ``Towards a Network of Atmospheric Cherenkov
  Detectors VII''}, 2005.

\bibitem{HESSModel3D}
M.~{Lemoine-Goumard et al.}
\newblock {\em APh}, 25:195, 2006.

\bibitem{slane}
P.~{Slane et al.}
\newblock {\em ApJ}, 548:614, 2001.

\bibitem{HESSCamera}
P.~{Vincent ({\it H.E.S.S. Collaboration})}.
\newblock In {\em Proceedings of the 28th ICRC, T. Kajita et al., Eds.
  (Universal Academy Press, Tokyo}, page 2887, 2003.

\end{thebibliography}
\bibliographystyle{plain}
\end{document}